# Distinct evolutions of Weyl fermion quasiparticles and Fermi arcs with bulk band topology in Weyl semimetals

N. Xu[1,2,§], G. Autès[2,3], C. E. Matt[1], B. Q. Lv[1,4], M. Y. Yao[1], F. Bisti[1], V. N. Strocov[1], D. Gawryluk[5], E. Pomjakushina[5], K. Conder[5], N. C. Plumb[1], M. Radovic[1], T. Qian[4], O. V. Yazyev[2,3] J. Mesot[1,2,6], H. Ding[5,7] and M. Shi[1,§]

[1] *Swiss Light Source, Paul Scherrer Institut, CH-5232 Villigen PSI, Switzerland*

[2] *Institute of Physics, École Polytechnique Fédérale de Lausanne (EPFL), CH-1015 Lausanne, Switzerland*

[3] *National Centre for Computational Design and Discovery of Novel Materials MARVEL, Ecole Polytechnique Fédérale de Lausanne (EPFL), CH-1015 Lausanne, Switzerland*

[4] *Beijing National Laboratory for Condensed Matter Physics and Institute of Physics, Chinese Academy of Sciences, Beijing 100190, China*

[5] *Laboratory for Developments and Methods, Paul Scherrer Institut, CH-5232 Villigen, Switzerland*

[6] *Laboratory for Solid State Physics, ETH Zürich, CH-8093 Zürich, Switzerland*

[7] *Collaborative Innovation Center of Quantum Matter, Beijing, 100190, China*

§ E-mail: nan.xu@psi.ch , ming.shi@psi.ch

The Weyl semimetal phase is a recently discovered topological quantum state of matter characterized by the presence of topologically protected degeneracies near the Fermi level. These degeneracies are the source of exotic phenomena, including the realization of chiral Weyl fermions as quasiparticles in the bulk and the formation of Fermi arc states on the surfaces. Here, we demonstrate that these two key signatures show distinct evolutions with the bulk band topology by performing angle-resolved photoemission spectroscopy, supported by first-principle calculations, on transition-metal monophosphides. While Weyl fermion quasiparticles exist only when the chemical potential is located between two saddle points of the Weyl cone features, the Fermi arc states extend in a larger energy scale and are robust across the bulk Lifshitz transitions associated with the recombination of two non-trivial Fermi surfaces enclosing one Weyl point into a single trivial Fermi surface enclosing two Weyl points of opposite chirality. Therefore, in some systems (e.g. NbP), topological Fermi arc states are preserved even if Weyl fermion quasiparticles are absent in the bulk. Our findings not only provide insight into the relationship between the exotic physical phenomena and the intrinsic bulk band topology in Weyl semimetals, but also resolve the apparent puzzle of the different magneto-transport properties observed in TaAs, TaP and NbP, where the Fermi arc states are similar.

Although any evidence of chiral Weyl fermions as fundamental particles in vacuum is still lacking, recent advances in topological condensed matter physics [1,2] provide a new way of realizing them in Weyl semimetals (WSMs). In an ideal WSM, bands without spin degeneracy cross at pairs of points at the Fermi level, the Weyl nodes, and the corresponding low-energy excitations fulfill the Weyl equation [3,4]. As a consequence, the quasiparticles in a WSM show the same exotic behaviors as Weyl fermions, such as the chiral anomaly effect [5-7] that can induce a negative longitudinal magneto-resistance (MR). Another hallmark of a WSM is the existence of anomalous surface states protected by the bulk band topology, the so-called Fermi arcs [8,9]. The WSM state has been proposed in magnetic systems [8, 10-19], Dirac semimetals [20-27] under magnetic field [28], as well as photonic crystals [29,30].

Recently, non-magnetic transition-metal monopnictides with broken inversion symmetry (the crystal structure and the first Brillouin zone (BZ) of these materials are shown in Fig. 1a-c) have been predicted to host the WSM phase [31,32]. Subsequently, pairs of Weyl cones have been experimentally observed in bulk TaAs [33,34], TaP [35,36] and NbAs [37]. On the other hand, although the Fermi arcs, which are one of the spectroscopic hallmarks of a WSM, are observed in all members of the transition-metal monopnictide family at both pnictogen- [34,36-40] and metal-terminated [35,41] surfaces, the magnetotransport behaviors related to the chiral anomaly effect are distinct among the different compounds. The negative longitudinal MR, which has a maximum value of -30% at ~6 T in TaAs [42] and as large as -3000% at 9T in TaP without any sign of saturation [43,44], has never been observed in NbP [45,46]. The apparent disagreement between the transport and spectroscopic signatures of WSM raises the questions of whether the observation of topological Fermi arcs is sufficient to identify a WSM phase [38,47], and whether the negative longitudinal MR in transition-metal monopnictides is not induced by the chiral anomaly but rather by the axial anomaly effect, which is not related to the Weyl fermion quasiparticles [48].

Here, we present a comparative study of transition-metal monopnictides by combining angle-resolved photoemission spectroscopy (ARPES) and first-principles calculations. We demonstrate that although the Weyl cones in NbP are very similar to those in TaAs and TaP, the chemical potential in NbP is located outside the energy window between two saddle points of the Weyl cone structures. Therefore, NbP is topologically different from the others with no Weyl fermion quasiparticle excitations being present. On the other hand, topological Fermi arc states in NbP and TaP disperse over larger energy scales, which are not directly related to the energies of bulk Lifshitz transitions. Our results not only provide a comprehensive understanding of the spectroscopic and magnetotransport signatures of transition-metal monopnictides, which can be extended to other WSMs, but also solve the apparent controversy of the experimental results in TaAs-family WSMs.

Single crystals of NbP (TaP) were grown by chemical vapor transport in a temperature gradient 850 °C→950 °C, using 0.6 g of polycrystalline NbP (TaP) as a source and iodine as a transport agent with a concentration of 12.2 mg/cm$^3$. Polycrystalline NbP (TaP) was synthesized by a solid-state reaction using elemental niobium (tantalum) and red phosphorus with a minimum purity of 99.99 %. The respective amounts of starting reagents were mixed and pressed into pellets in a He-glove box and annealed in an evacuated quartz ampule at 1050 °C for 60 hours. Laboratory x-ray diffraction measurements, which were done at room temperature using Cu Ka radiation on a Brucker D8 diffractometer, have shown that the obtained crystals are single phase with the tetragonal structure of space group I41/amd (N 141). Clean surfaces for ARPES measurements were obtained by cleaving NbP and TaP samples in a vacuum better than $5 \times 10^{-11}$ Torr. Surface-sensitive VUV-ARPES measurements were performed with circular-polarized light at the Surface/Interface Spectroscopy (SIS) beamline at the Swiss Light Source (SLS) with a Scienta R4000. Bulk-sensitive soft X-ray ARPES measurements were performed at the Advanced Resonant Spectroscopies (ADRESS) beamline at SLS [49] using a SPECS analyser, and data were collected using circular-polarized light with an overall energy resolution in the range of 50-80 meV at $T = 10$ K. The electronic structure of NbP was computed within the density functional theory (DFT) framework using the generalized gradient approximation (GGA) as implemented in the QUANTUM-ESPRESSO software package [50]. Spin-orbit coupling (SOC) is taken into account with the help of relativistic pseudopotentials [51]. The calculations were carried out using an 8×8×8 *k*-point mesh and a planewave kinetic energy cutoff of 50 Ry for the wavefunctions. We used the experimentally determined crystal structure from Ref. [52]. To compute the surface density of states, a tight-binding Hamiltonian was derived in the bulk Wannier functions basis for Nb *d*- and P *p*-orbitals [53] and the surface Green's function of the semi-infinite system was obtained according to the method introduced in Ref. [54]. Possible effects resulting from the surface relaxation were neglected.

First, we present our first-principles calculations performed on NbP (Figs. 1d-g). In agreement with previous theoretical studies [10-11], we find that two types of Weyl nodes exist in NbP, eight pairs of W1 type (Fig. 1d) located away from the $k_z = 0$ plane and four pairs of W2 type (Fig. 1e) located in the $k_z = 0$ plane. Figure 1f shows the calculated surface local density of states, while Figs. 1g,h provide a magnified view near the surface projections of W1 nodes. The Fermi arc states at the P-termination (a natural cleavage plane indicated in Fig. 1b) are clearly separated from both the trivial surface states and the surface projection of the bulk bands. However, the end points of the Fermi arc do not exactly connect to the projections of W1 nodes on the surface BZ, but instead show a small separation. This is different from TaAs and TaP [33-36, 39], in which the Fermi arcs terminate at the exact locations of W1 nodes projected onto the surface BZ. We will show that this distinction is related to the absence of Weyl fermion quasiparticles in NbP, which makes it topologically different from TaAs and TaP.

To verify the calculated bulk states experimentally, we performed soft x-ray ARPES (SX-ARPES) measurements of NbP in order to achieve higher bulk sensitivity [49, 55]. The bulk origin of the probed electronic states is confirmed by the three-dimensional Fermi surface (FS) seen in the *out-of-plane* FS map (Fig. 2a), taken in the *hν* range of 400–600 eV. The determined periodicity of π/*c'* (*c'* = *c/2*) in the *out-of-plane* direction is fully consistent with the crystal structure of NbP. The FS contours in $k_z$ are clearly resolved in the SX-ARPES energy range due to the increase of the photoelectron mean free path [56]. The observed band structure along high symmetry path Γ-Σ-S-Z-Γ (Fig. 2b) agrees well with the bulk states from calculations with spin-orbit coupling included (red lines in Fig. 2b). In Fig. 2c-e, we plot the ARPES intensity along momentum cuts passing through the location of theoretically predicted W1 node, (0.54 π/*a*, 0.01 π/*a*, 0.57 π/*c'*), in the directions of $k_x$, $k_y$ and $k_z$, respectively. In agreement with the calculations, only a hole-like pocket is observed, which corresponds to the bottom part of the predicted Weyl cone, since the W1 node is located far above the Fermi level in NbP. On the other hand, both the upper and lower parts of the Weyl cone structure of W2 are clearly observed at $k_x = 1.02$ π/*a* in the $k_z = 0$ plane (Fig. 2f-h), which is in good agreement with the calculations.

Compared to TaAs and TaP [33-36], bulk states in NbP show a distinct feature, namely that the energies of Weyl nodes W1 and W2 ($E_{W1}$ and $E_{W2}$) are relatively far above/below the chemical potential (Fig. 2c-h). As a consequence, the pairs of W1 (W2) nodes are enclosed by single valence (conduction) FS pockets with a zero net FS Chern number ($C_{FS} = 0$) [57-59]. Such a Fermi surface is topologically trivial and chiral excitations cannot occur. Therefore, our results indicate that, in contrast to TaAs and TaP, in NbP there are no low-energy excitations near $E_F$ that could correspond to the Weyl fermion quasiparticles. This provides a natural explanation for the absence of negative longitudinal MR in this monopnictide compound [45,46].

The absence of Weyl fermion quasiparticles in NbP provides an opportunity to study the evolution of the Fermi arc states with respect to the bulk FS topology. Figures 3a-c show the FS map, band structure along $\bar{\Gamma}$-$\bar{Y}$ direction and the corresponding curvature plot of the surface states of NbP probed by ARPES using photon energy $hv = 50$ eV to enhance the surface sensitivity. In agreement with previous studies of the surface states in NbP [40-41], we observed the Fermi arc states that dominate the ARPES signal and appear near the surface projections of the W1 Weyl nodes (black dashed line area in Fig. 3a). A trivial surface band has also been observed. Although the arc states and the trivial surface band are very close at the Fermi level, their contributions can be resolved at higher binding energies (Fig. 3b-c). The Fermi arcs in TaAs and TaP connect to the surface projections of W1 Weyl nodes, as observed by the ARPES experiments [33-36,38]. In contrast, Fermi arc states observed in NbP do not connect to the surface projections of W1 nodes (blue dots in Fig. 3d) at the Fermi energy. The separation of the end points of the Fermi arc is twice as large as the calculated distance between the nodes. A consistent picture of the Fermi arc states is also suggested by the surface local density of states calculated for the P-termination shown in Fig. 3h. Shifting the energy to $E_{W1}$ (as indicated by the bulk band calculations in Fig. 3e), the length of the Fermi arc shrinks and their end points terminate at the surface projections of the W1 nodes, as illustrated by the calculated constant energy map shown in Fig. 3g. A further shift of the chemical potential to the energy where the pair of W1 nodes is enclosed by a single trivial FS does not lead to a disappearance of the arcs. Instead the Fermi arcs merge into the bulk Weyl cones (Fig. 3f). A similar Fermi arc evolution was also observed in TaP.

The Fermi arc states in TaP connect to the projections of W1 nodes at the Fermi level (Fig. 3i), where its energy is naturally close to $E_{W1}$ [35]. After shifting the chemical potential 150 meV downwards, the Weyl fermion quasiparticles vanish since the bulk FSs enclosing W1 nodes go through a transition from two topological FSs ($C_{FS}= \pm 1$) to a trivial one ($C_{FS}= \pm 0$) (Fig. 3k), but the Fermi arc states are still present and extend in momentum space (Fig. 3j). Our results for TaP and NbP reveal a universal evolution of the Fermi arc surface states, and demonstrate that these states disperse over a much larger energy range, compared to the energy range of the existence of Weyl fermion quasiparticles. For this reason, Fermi arcs can be observed in compounds such as NbP [40,41], even when there are no Weyl fermion quasiparticles and associated magnetotransport phenomena [45,46].

Our results are summarized in Figure 4. First, the topological nature of NbP is different from that of TaAs and TaP. As illustrated in Fig. 4a, the chiral FSs ($C_{FS}= \pm 1$) with low-energy excitations corresponding to the Weyl fermion quasiparticles exist only when the chemical potential is located in the energy window between two saddle points of the Weyl cone structure ($E_{SP1}$ and $E_{SP2}$). In contrast to TaP [35] and TaAs [33] where one or two types of Weyl fermions contribute to the magnetotransport properties, respectively, there are no Weyl fermion quasiparticles in pristine NbP due to the positions of $E_{SP1}$ and $E_{SP2}$ relative to the Fermi level (Fig. 4b). This explains the absence of the transport signature of chiral anomaly in NbP [45,46]. On the other hand, the topological Fermi arc surface states disperse in a larger energy range than the Weyl fermion quasiparticles. As sketched in Figs. 4c-d, Fermi arc surface states are not sensitive to the bulk Lifshitz transitions in NbP and TaP. Even for a system without any chiral FSs and Weyl fermion quasiparticles (as $E_F > E_{SP1}$ or $E_F < E_{SP2}$ in NbP), Fermi arcs can still be observed by spectroscopic techniques. In this case, the Fermi arc surface states are no longer connecting the projections of the Weyl nodes, but instead terminate in the bulk Weyl cones. Therefore, although the physical properties related to the Weyl fermion quasiparticles are absent, phenomena associated with the topological Fermi arc surface states are still preserved and can be detected experimentally. This distinct evolution of Weyl fermion quasiparticles and Fermi arcs with respect to the bulk band topology is present in TaP and NbP, and could be extended to other WSMs. Our findings not only provide a comprehensive understanding of the relationship between the exotic physical phenomena and the intrinsic bulk Fermi surface topology in Weyl semimetals, but also reconcile the controversy between the transport and spectroscopy experiments in NbP.

## Acknowledgements

N.X. and M.S. acknowledge the support by the Sino-Swiss Science and Technology Cooperation (No. IZLCZ2138954). This work was supported by NCCR-MARVEL funded by the Swiss National Science Foundation, the Swiss National Science Foundation (No. 200021-137783), the Ministry of Science and Technology of China (No. 2013CB921700, No. 2015CB921300, No. 2011CBA00108 and No. 2011CBA001000), the National Natural Science Foundation of China (No. 11474340, No. 11422428, No. 11274362 and No. 11234014), the Chinese Academy of Sciences (No. XDB07000000). G. A. and O. V. Y. acknowledge support by the NCCR Marvel and the ERC Starting grant "TopoMat" (Grant No. 306504). First-principles electronic structure calculations have been performed at the Swiss National Supercomputing Centre (CSCS) under Project No. s675. F.B. acknowledges funding from the Swiss National Science Foundation under the grant 200021_146890 and from the European Community's Seventh Framework Program (FP7/2007-2013) under the grant 290605 (PSI FELLOW/COFUND).

## References

[1]  M. Z. Hasan and C. L. Kane, Rev. Mod. Phys. **82**, 3045 (2010).

[2]  X. L. Qi and S. C. Zhang, Rev. Mod. Phys. **83**, 1057 (2011).

[3]  H. Weyl, Z. Phys. **56**, 330– 352 (1929).

[4]  H. B. Nielsen and Masao Ninomiya, Physics Letters B **130**, 389 (1983).

[5]  H. B. Nielsen and Masao Ninomiya, Physics Letters B **130**, 389 (1983).

[6]  D. T. Son and B. Z. Spivak, Phys. Rev. B **88**, 104412 (2013).

[7]  Pavan Hosur and Xiaoliang Qi, Comptes Rendus Physique **14**, 857 (2013).

[8]  Xiangang Wan, Ari M. Turner, Ashvin Vishwanath, and Sergey Y. Savrasov, Phys. Rev. B **83**, 205101 (2011).

[9]  Leon Balents, Physics **4**, 36 (2011).

[10] Gang Xu, Hongming Weng, Zhijun Wang, Xi Dai, and Zhong Fang, Phys. Rev. Lett. **107**, 186806 (2011).

[11] Daniel Bulmash, Chao-Xing Liu, and Xiao-Liang Qi, Phys. Rev. B **89**, 081106 (2014).

[12] A. A. Burkov and Leon Balents, Phys. Rev. Lett. **107**, 127205 (2011).

[13] Gábor B. Halász and Leon Balents, Phys. Rev. B **85**, 035103 (2012).

[14] A. A. Zyuzin, Si Wu, and A. A. Burkov, Phys. Rev. B **85**, 165110 (2012).

[15] Tena Dubcek, Colin J. Kennedy, Ling Lu, Wolfgang Ketterle, Marin Soljačić, Hrvoje Buljan, Phys. Rev. Lett., **114**, 225301 (2015).

[16] M. Hirayama, R. Okugawa, S. Ishibashi, S. Murakami, and T. Miyake, Phys. Rev. Lett. **114**, 206401 (2015).

[17] Jianpeng Liu and David Vanderbilt, Phys. Rev. B **90**, 155316 (2014).

[18] Bahadur Singh, Ashutosh Sharma, H. Lin, M. Z. Hasan, R. Prasad, and A. Bansil, Phys. Rev. B **86**, 115208 (2012)

[19] Tomas Bzdusek, Andreas Ruegg, Manfred Sigrist, Phys. Rev. B **91**, 165105 (2015).

[20] Zhijun Wang, Yan Sun, Xing-Qiu Chen, Cesare Franchini, Gang Xu, Hongming Weng, Xi Dai, and Zhong Fang, Phys. Rev. B **85**, 195320 (2012).

[21] Zhijun Wang, Hongming Weng, Quansheng Wu, Xi Dai, and Zhong Fang, Phys. Rev. B **88**, 125427 (2013).

[22] Z. K. Liu, B. Zhou, Y. Zhang, Z. J. Wang, H. M. Weng, D. Prabhakaran, S.-K. Mo, Z. X. Shen, Z. Fang, X. Dai, Z. Hussain, and Y. L. Chen, Science **343**, 864 (2014).

[23] Z. K. Liu, J. Jiang, B. Zhou, Z. J. Wang, Y. Zhang, H. M. Weng, D. Prabhakaran, S. K. Mo, H. Peng, P. Dudin, T. Kim, M. Hoesch, Z. Fang, X. Dai, Z. X. Shen, D. L. Feng, Z. Hussain, and Y. L. Chen, Nat. Mater. **13**, 677 (2014).

[24] M. Neupane, S.-Y. Xu, R. Sankar, N. Alidoust, G. Bian, C. Liu, I. Belopolski, T.-R. Chang, H.-T. Jeng, H. Lin, A. Bansil, F. Chou, and M. Z. Hasan, Nat. Commun. **5**, 3786 (2014).

[25] S. Borisenko, Q. Gibson, D. Evtushinsky, V. Zabolotnyy, B. Buchner, and R. J. Cava, Physical Review Letters **113**, 027603 (2014).

[26] Hongming Weng, Yunye Liang, Qiunan Xu, Rui Yu, Zhong Fang, Xi Dai, and Yoshiyuki Kawazoe, Phys. Rev. B **92**, 045108 (2015).

[27] Rui Yu, Hongming Weng, Zhong Fang, Xi Dai, Xiao Hu, Phys. Rev. Lett. **115**, 036807 (2015).

[28] J. Xiong *et al*. Science **350**, 413 (2015).

[29] Ling Lu, Liang Fu, John D Joannopoulos, and Marin Soljačić, Nature Photonics **7**, 294 (2013).

[30] Ling Lu, Zhiyu Wang, Dexin Ye, Lixin Ran, Liang Fu, John D. Joannopoulos, and Marin Soljačić, Science **349**, 622, (2015).

[31] Hongming Weng, Chen Fang, Zhong Fang, B. A. Bernevig and X. Dai, Phys. Rev. X **5**, 011029 (2015).

[32] S. M. Huang *et al*. Nature Commun. **6**:7373 (2015)

[33] B. Q. Lv, N. Xu, H. M. Weng, J. Z. Ma, P. Richard, X. C. Huang, L. X. Zhao, G. F. Chen, C. Matt, F. Bisti, V. N. Strocov, J. Mesot, Z. Fang, X. Dai, T. Qian, M. Shi, and H. Ding, Nature Physics **11**, 724-727 (2015).

[34] S. Y. Xu *et al.*, Science **349**, 613 (2015)

[35] N. Xu, H. M. Weng, B. Q. Lv, C. Matt, J. Park, F. Bisti, V. N. Strocov, E. Pomjakushina, K. Conder, N. C. Plumb, M. Radovic, G. Autès, O. V. Yazyev, Z. Fang, X. Dai, G. Aeppli, T. Qian, J. Mesot, H. Ding, M. Shi, Nat. Commun. **7**, 11006 (2016).

[36] S.-Y. Xu, I. Belopolski, D. S. Sanchez *et al.*, Science Advances **1**, 1501092 (2015).

[37] S. Y. Xu *et al.*, Nature Physics **11**, 748 (2015).

[38] B. Q. Lv, H. M. Weng, B. B. Fu, X. P. Wang, H. Miao, J. Ma, P. Richard, X. C. Huang, L. X. Zhao, G. F. Chen, Z. Fang, X. Dai, T. Qian, H. Ding, Phys. Rev. X **5**, 031013 (2015).

[39] L. X. Yang *et al.*, Nature Physics **11**, 728–732 (2015).

[40] D. F. Xu *et al.*, Chin. Phys. Lett. **32**, 107101 (2015)

[41] S. Souma, Z. Wang, H. Kotaka, T. Sato, K. Nakayama, Y. Tanaka, H. Kimizuka, T. Takahashi, K. Yamauchi, T. Oguchi, K. Segawa, and Y. Ando, Phys. Rev. B **93**, 161112 (2016).

[42] X. Huang, L. Zhao, Y. Long, P. Wang, D. Chen, Z. Yang, H. Liang, M. Xue, Hongming Weng, Z. Fang, X. Dai, G. Chen, Phys. Rev. X **5**, 031023 (2015).

[43] C. Shekhar, F. Arnold, S. Wu, Y. Sun, M. Schmidt, N. Kumar, A. Grushin, J. H. Bardarson, R. Reis, M. Naumann, M. Baenitz, H. Borrmann, M. Nicklas, E. Hassinger, C. Felser, and B. Yan, Nat. Commun. **7**:11615 (2016).

[44] Jianhua Du *et al.*, Sci. China-Phys. Mech. Astron. **59**, 657406 (2016).

[45] C. Shekhar, A. K. Nayak, Y. Sun, M. Schmidt, M. Nicklas, I. Leermakers, U. Zeitler, W. Schnelle, J. Grin, C. Felser and B. Yan, Nature Physics **11**, 645 (2015).

[46] Chenglong Zhang *et al.*, Phys. Rev. B **92**, 041203(R) (2015).

[47] Ilya Belopolski *et al.*, Phys. Rev. Lett. **116**, 066802 (2016).

[48] Pallab Goswami, J. H. Pixley, and S. Das Sarma, Phys. Rev. B **92**, 075205 (2015).

[49] V. N. Strocov *et al.*, J. Synchrotron Rad. **21**, 32–44 (2014).

[50] P. Giannozzi et al., J. Phys. Condens. Matter 21, 395502 (2009).

[51] A. Dal Corso and A. Mosca Conte, Phys. Rev. B 71, 115106 (2005).

[52] J. O. Willerstroem, Journal of the Less-Common Metals 99, p273-p283 (984).

[53] A. A. Mostofi, J. R. Yates, G. Pizzi, Y. S. Lee, I. Souza, D. Vanderbilt, N. Marzari, Comput. Phys. Commun. 185, 2309 (2014).

[54] A. Umerski, Phys. Rev. B 55, 5266 (1997).

[55] V. N. Strocov, J. Electron Spectrosc. Relat. Phenom. **130**, 65−78 (2003).

[56] V.N. Strocov, M. Shi, M. Kobayashi, C. Monney, X. Wang, J. Krempasky, T. Schmitt, L. Patthey, H. Berger & P. Blaha, Phys. Rev. Lett. **109**, 086401 (2012).

[57] Z. Fang, N. Nagaosa, K. S. Takahashi, A. Asamitsu, R. Mathieu, T. Ogasawara, H. Yamada, M. Kawasaki, Y. Tokura, and K. Terakura, Science **302**, 92 (2003).

[58] Hongming Weng, Rui Yu, Xiao Hu, Xi Dai and Zhong Fang, Adv. Phys. **64**, 227 (2015).

[59] P. Horava, Phys. Rev. Lett. **95**, 016405 (2005).

# Figures

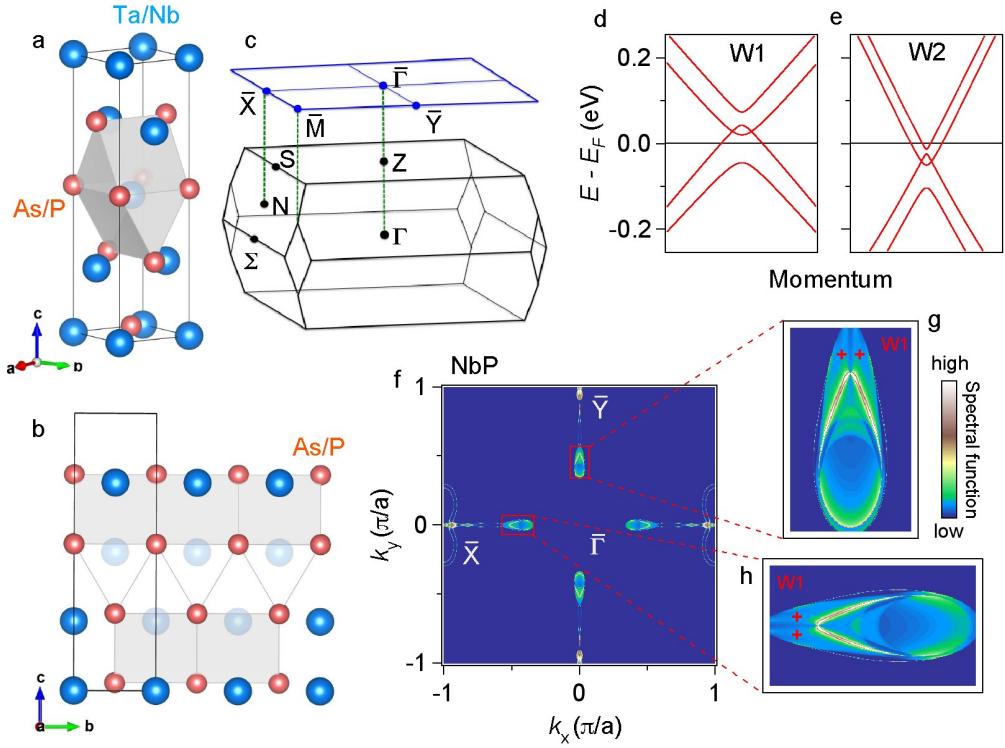

**Figure 1. Predicted bulk Weyl cones and surface-state Fermi arcs in transition-metal monopnictides. a**, Crystal structure of transition-metal monopnictides and the structure of **b**, As/P surface termination. **c**, Bulk and surface Brillouin zone with high-symmetry points labelled. **d-e**, Calculated bulk band structure of NbP along momentum directions passing through pairs of Weyl nodes W1 and W2, respectively. **f**, Calculated momentum-resolved surface local density of states for the P-termination of NbP at the Fermi level and **g-h**, the magnified view of the Fermi arc states near the W1 node along the $\bar{\Gamma}$-$\bar{Y}$ and $\bar{\Gamma}$-$\bar{X}$ directions, respectively.

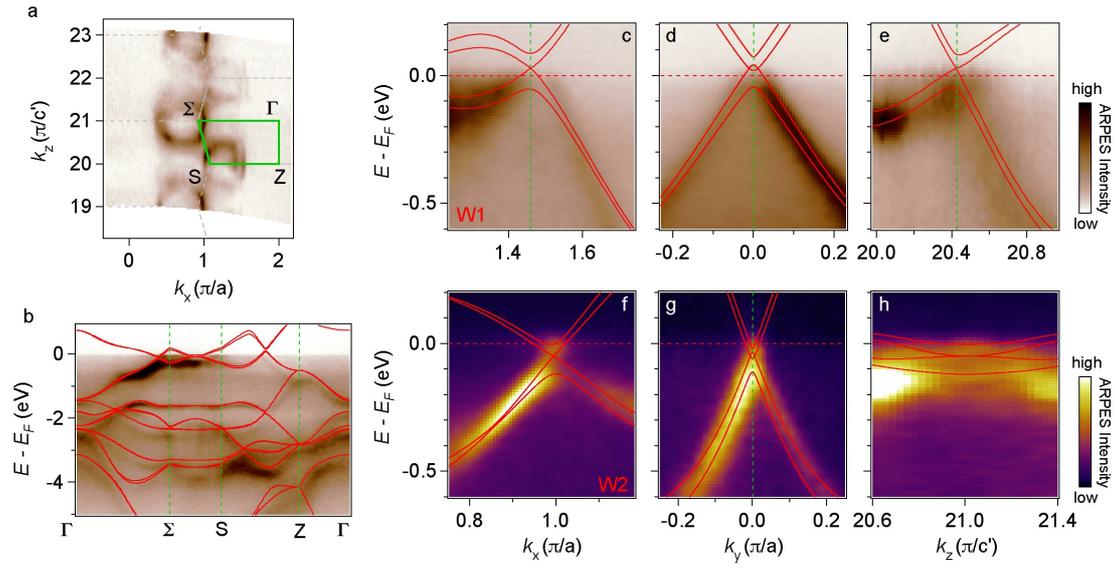

**Figure 2. Experimentally observed bulk electronic structure and Weyl cones in NbP. a-b**, The out-of-plane FS and band dispersions along high symmetry lines in NbP acquired with SX-ARPES. **c-e**, Photoemission intensity plot along momentum cuts passing through W1 in the $k_x$, $k_y$, and $k_z$ directions, respectively. **f-h**, Same as **c-e** but for W2. The overlaid lines are the calculated bands for comparison.

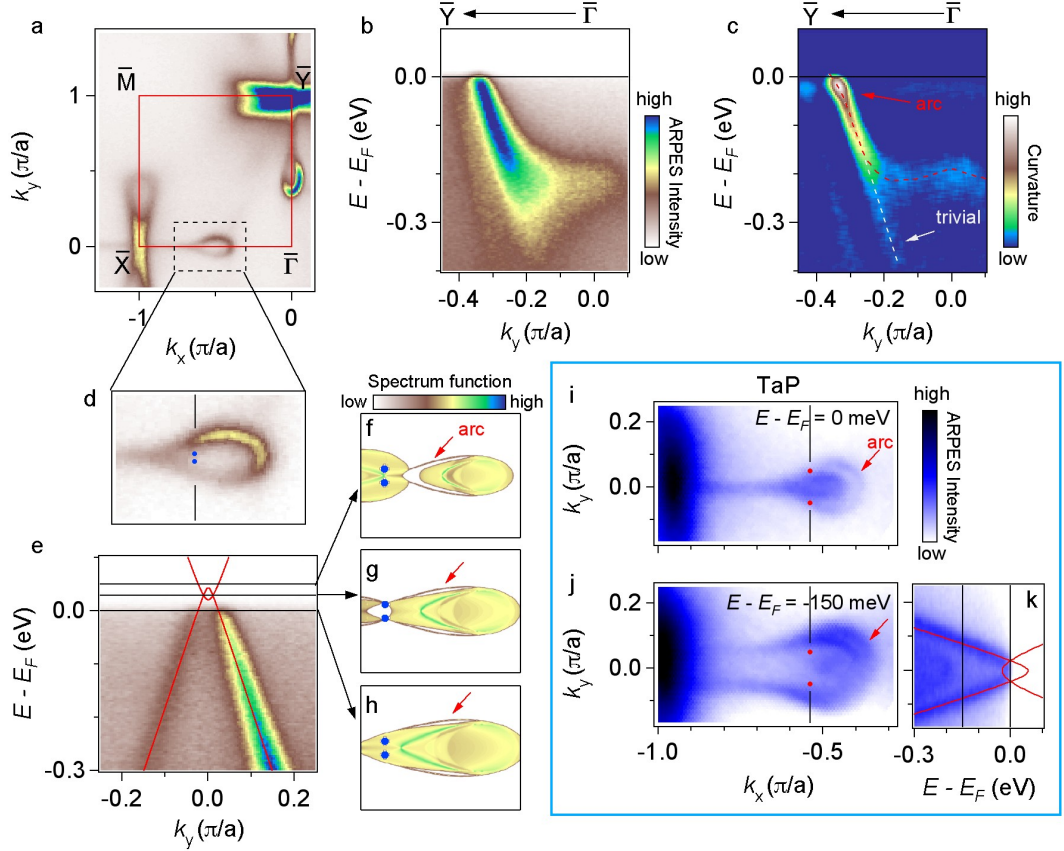

**Figure 3. Fermi arc states at the P-terminated (001) surface of NbP and TaP. a**, The FS intensity map of NbP acquired by ARPES with $h\nu$ = 50 eV. **b-c**, The ARPES spectrum and the corresponding curvature plots along the $\bar{\Gamma}$-$\bar{Y}$ direction, respectively. **d**, Zoomed-in FS. The blue dots indicate the projections of the W1 nodes extracted from DFT calculations. **e**, Surface states along a momentum cut passing through pairs of the projections of W1 nodes. The bulk bands passing through pairs of W1 nodes are overlaid for comparison. **f-h**, Calculated surface states at the P-terminated (001) surface for various binding energies, as indicated in **e**. **i-j**, ARPES intensity map on TaP at constant energies the Fermi level and 150 meV below it, acquired with $h\nu$ = 50 eV. The projections of W1 nodes are indicated by the red dots. **k**, The surface states along a momentum cut passing through a pair of projections of W1 nodes, with the calculated bulk band passing through the pair of W1 nodes overlaid for comparison.

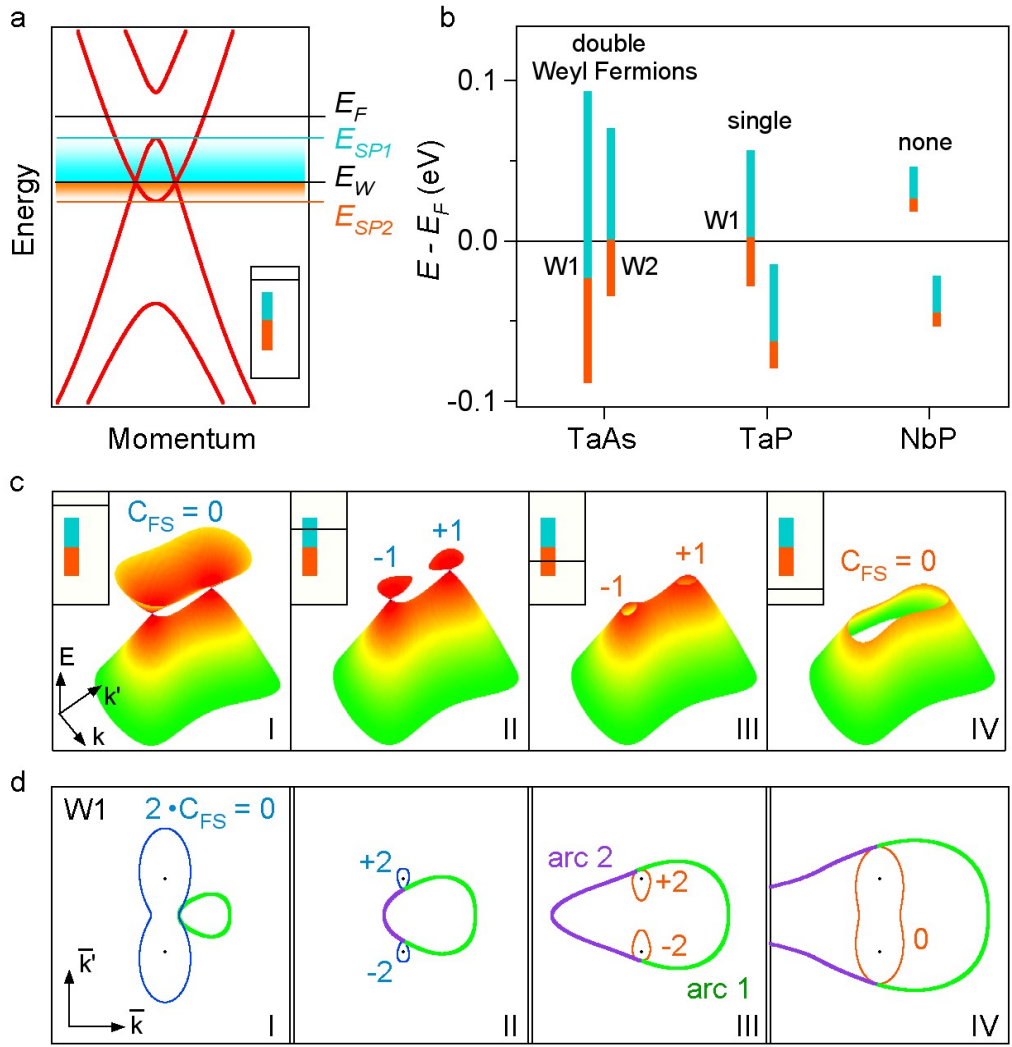

**Figure 4. Evolutions of Weyl fermion quasiparticles and the Fermi arc surface states in transition-metal monopnictides. a**, Band structure along a momentum cut containing a pair of Weyl nodes, the critical energy scales are indicated by horizontal lines. The inset shows the simplified sketch of the energy positions. **b**, Diagram of the relationship between the energy windows of the Weyl fermion quasiparticles and chemical potentials in TaAs, TaP and NbP. **c**, Illustrations of the FS topology and Fermi surface Chern number evolutions with chemical potentials in a WSM. **d**, Corresponding evolution of the Fermi arc surface states in TaP and NbP.